\documentclass[prl,aps,twocolumn,floats]{revtex4}

\usepackage{graphicx}

\begin{document}
\unitlength 1 cm
\newcommand{\be}{\begin{equation}}
\newcommand{\ee}{\end{equation}}
\newcommand{\bearr}{\begin{eqnarray}}
\newcommand{\eearr}{\end{eqnarray}}
\newcommand{\nn}{\nonumber}
\newcommand{\vk}{\vec k}
\newcommand{\vp}{\vec p}
\newcommand{\vq}{\vec q}
\newcommand{\vkp}{\vec {k'}}
\newcommand{\vpp}{\vec {p'}}
\newcommand{\vqp}{\vec {q'}}
\newcommand{\bk}{{\bf k}}
\newcommand{\bp}{{\bf p}}
\newcommand{\bq}{{\bf q}}
\newcommand{\br}{{\bf r}}
\newcommand{\up}{\uparrow}
\newcommand{\down}{\downarrow}
\newcommand{\fns}{\footnotesize}
\newcommand{\ns}{\normalsize}
\newcommand{\cdag}{c^{\dagger}}

\title{Theory of Neutron Scattering for\\ 
Gapless Neutral Spin-1 Collective Mode in Graphite}
\author{S. A. Jafari$^{1,2}$ and G. Baskaran$^2$}
\address{$^1$ Department of Physics, Sharif University of Technology, 
Tehran 11365-9161, Iran\\
{$^2$}Institute of Mathematical Sciences, Madras 600 113, India\\ }


\begin{abstract}
Using tight binding band picture for 2D graphite, and the Hubbard 
interaction, recently we obtained a gapless, neutral spin-1 collective 
mode in graphite \cite{SZS}. In this paper we present a detailed RPA 
analysis of the Neutron Scattering cross section for this collective mode. 
Near $K-$point and very close to $\Gamma-$point, the intensity of neutron 
scattering peaks vanishes as $q^3$. This is shown using a simple Dirac
cone model for the graphite band structure, which captures the small$-q$ 
behavior of the system. As we move away from the $\Gamma-$ and $K-$points 
in the Brillouin zone of the collective mode momenta, we can identify our 
collective mode quanta with spin triplet excitons with the spatial extent 
of the order of a few to a couple of lattice parameter $a$, with more or 
less anisotropic character, which differs from point to point. We also 
demonstrate that the inclusion of the long range tail of the Coulomb 
interaction in real graphite, does not affect our spin-1 collective mode 
qualitatively.  This collective mode could be probed at different energy 
scales by thermal, hot and epithermal neutron scattering experiments.
However, the smallness of the calculated scattering intensity, arising
from a reduced form factor of carbon $2p_z$ orbital makes the detection 
challenging. 
\end{abstract}

\maketitle  

\section{Introduction}
Graphite is a broad band tight-binding system composed of hexagonal graphene 
sheets held together by van der Waals interaction. This pure carbon system,
inspite of a simplicity in electronic and crystal structure, has a rich 
physics and continues to surprise us\cite{kopelevich}.  
The four valence of each 
carbon atom is used to form $\sigma$ and $\pi$ bonds with its three neighbors. 
Each layer is like a giant molecule with {\em resonating} $\pi$ bonds
among many valence-bond configurations \cite{Pauling}. This structure is 
similar to many interesting systems such as carbon nano-tubes, buckyball, 
$MgB_2$ etc.  Within each graphene layer, $p_z$ electrons of each carbon 
atom can hop from a site to another site. These electrons are responsible 
for the formation of $\pi$ bands which touch each other at the corners 
of the Brillouin zone (Figure \ref{vband.fig})

  An important question is to understand the nature of low energy 
collective excitations in our zero gap planar $p-\pi$ bonded graphite. 
We may approach this question in an unconventional fashion starting from 
organic chemistry ! Planar $p-\pi$ bonded molecules form the basis of 
organic chemistry, 
benzene being a first member. Benzene is a mini-graphite in some respects. 
One can also view graphite as an end member of planar $p-\pi$ bonded 
molecules - benzene, naphthalene, anthracene, coronene etc. It is well 
known\cite{Jortner} that benzene and the above sequence have a spin triplet 
state as their first excited state. The next excited state is a singlet 
state nearly $2~eV$ above the triplet state for benzene. This remarkable 
singlet-triplet splitting, which is missing in simple Huckel theory, is a 
well known effect of coulomb correlation in $p-\pi$ bonded systems. 

A natural question is what happens to the triplet and singlet excitons 
in graphite, the end member of the above sequence. In a recent paper 
we showed that in graphite
the triplet excitons form a well defined band in the entire Brillouin 
zone. We view the low energy part of the above exciton as a spin-1 
collective mode in view of its gapless character. The singlet excitons,
on the other hand form the well known $\pi$ plasmon (energy $\sim 7~eV$) 
with a finite gap in the spectrum. Since graphite is also viewed as a 
semi-metal, our neutral spin-1 collective mode can be also viewed as 
Landau's `Spin-1 Zero Sound'(SZS). 

Another way to understand the low energy collective mode in graphite
is to go back to Pauling's resonating valence bond (RVB) theory of 
graphite\cite{Pauling}. Pauling assumed a dominant near neighbor singlet 
$p-\pi$ bonds and their resonance to develop a theory of graphite. 
We know from recent developments in RVB theory that such a well developed 
singlet correlation leads to a quantum spin liquid state. A quantum spin 
liquid state either has a collective spin-1 branch or contain spin-half
spinon excitations by quantum number fractionization. Our finding of a 
gapless spin-1 mode qualifies graphite to be a quantum spin liquid state.  
The gapless character of the spin-1 branch, however, makes it a long 
range RVB state rather than a short range RVB state as envisaged by 
Pauling.

Yet another physical picture of our spin-1 collective mode is as follows. 
In graphite the valence band is completely filled and the conduction 
band is completely empty. The nature of particle-hole continuum in this 
system is such that a window below the particle-hole continuum opens (see 
figure \ref{ph_continuum.fig}). A Hubbard type on site repulsion between the 
electrons is in the spin singlet channel. This translates into an attraction 
in particle-hole spin triplet channel. This can bind an electron and
a hole into a spin-1 exciton; this appears as a pole for the spin 
susceptibility inside the window which exists below the particle-hole 
continuum. There are special band structure reasons as to why our spin-1 
excitations are gapless, which will be explained in detail below.
\begin{figure}[ht]
\begin{center}
\includegraphics[scale=0.40]{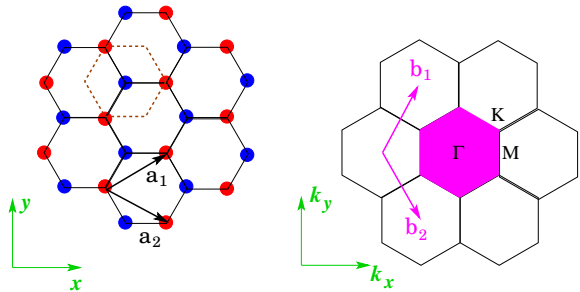}
\caption{(left) Lattice structure of a two-dimensional graphene sheets. It is a bipartite lattice composed of two sub-lattices. Each sub-lattice is denoted by a different color. The basis vectors of real space lattice are given by ${\vec a_1}=\frac{\sqrt 3 a}{2} \hat{e}_x+\frac{a}{2} \hat{e}_y,~{\vec a_2}=\frac{\sqrt 3 a}{2} \hat{e}_x-\frac{a}{2} \hat{e}_y$, where $\frac{a}{\sqrt 3}$ is nearest neighbor C-C distance. The unit cell is denoted by a dotted hexagon and contains two non-equivalent carbon atoms, belonging to two sub-lattices. (right) Corresponding reciprocal space is defined by ${\vec b_1}=\frac{2\pi}{a}(\frac{\hat{e}_x}{\sqrt 3}+\hat{e}_y),~  {\vec b_2}=\frac{2\pi}{a}(\frac{\hat{e}_x}{\sqrt 3}-\hat{e}_y)$. One possible choice for Brillouin zone (BZ) is a hexagonal region of the above figure.}
\label{lattice.fig}
\end{center}
\end{figure}
  Use of the short range interaction for the spin phenomena is justified, 
 since (as in the following we will explicitly demonstrate for the case of 
spin-1 collective mode) inclusion of the long range tail of the interaction 
does not affect the spin physics qualitatively. 

Gapless spin-1 excitation (Goldstone modes) are characteristic of 
magnetically ordered systems.  In the now popular case of $La_2CuO_4$, 
a 2D quantum antiferromagnet, the spin-wave bandwidth is of the order 
of $\sim 100~meV$.  But in the case of graphite, which has no long range 
magnetic order, this gapless spin-1 collective branch (a non-Goldstone mode) 
has a wide dispersion 
$\sim 0-2~eV$ (figure \ref{collective.fig}). This energy range should 
be compared with triplet exciton of buckyball at 
$\sim 1.64~eV$\cite{Goldoni}. Spin-1 collective mode of graphite exists 
everywhere in BZ and since it does not enter the particle-hole continuum, 
is protected form Landau damping (decaying into particle-hole pairs). 
Therefore the spin-1 collective mode of graphite is long-lived and could 
be exploited in coherent transport of {\em spin-only} (neutral) currents 
through the nano-tubes. 

 Since the bandwidth of this collective mode is very large, each energy 
range should be investigated by different probe. Here we will concentrate 
on inelastic neutron scattering experiments. In real graphite, there are 
very small electron-hole pockets around the $K-$points arising form 
inter-layer hopping of the order of $\sim 10-20~meV$. Above this energy 
range, the collective mode is well isolated from the typical phonon 
energies and therefore the neutrons need not to be polarized. The low 
energy parts (regions close to $\Gamma$-point in figure 
\ref{collective.fig}) of the collective mode up to $\sim 100~meV$ 
can be probed with thermal neutrons. Hot neutrons can concentrate on 
higher energies $\sim 100-500~meV$ (around the $\Gamma$-point). The 
highest energy parts $\sim 0.5~eV-$up-wards (regions midway between 
$\Gamma-M$ and midway between $\Gamma-K$) correspond to bound state wave 
function with spatial extent of the order of a few unit cell 
($\sim 2 \AA$) and can be studied by hot and/or epithermal neutrons.

However, an important practical difficulty is the small value of 
scattering cross section. As we will see in detail, the large size of
the spin carrying $2p_z$ orbital of carbon reduces the neutron scattering
form factor significantly, making $S({\bf q},\omega)$ rather small 
in relevant regions in $({\bf q},\omega)$ space. A most appropriate
region, where experiment have better chance for discovering the 
spin-1 mode is in the region midway between the $\Gamma$ and $M$ points
in the Brilluouin zone.

  In this paper we present the detailed RPA analysis of the spin 
susceptibility and the neutron scattering cross section. The bare value 
of Hubbard $U$ in graphite is $\sim 8~eV$, but the renormalized value 
for stability issues, should be less than $2.23 t\sim 5.8~eV$. 
We will keep $U$ as a parameter to be determined by experiment and 
will report the results of cross section calculation for $U\sim 5~eV$. 
The organization of the paper is as follows. To be self-contained, 
we begin with a brief review of the cross section of a neutron scattering 
experiment which is followed by a summaries of single particle band 
picture of the graphene which is essential for understanding the 
nature of the window below the particle-hole continuum. The RPA 
formulation of the next section is applied to tight-binding bands of 
graphene in later sections. To get an analytical handle, we exactly 
obtain the spin susceptibility within the RPA for {\em model of 
a single Dirac cone}. This model captures the behavior near the 
$\Gamma$ and $K-$point. We calculate the intensities of the neutron 
scattering peaks for the linearized model and also discuss the effect 
of including long range tail of the interaction. Finally we report 
the numerical calculation of the intensities of the neutron 
scattering peaks for entire BZ along with calculation of the real 
space profile of the excitonic wave-function. To be self-contained, 
in appendix A, we review excitonic formulation of the problem.

\section{Neutron Scattering cross section \newline Basic Formulation}
\label{NeutronScatteringFormulation}
  Neutron scattering is a very weak probe and does not disturb the target very much. It makes the linear response theory a very useful tool to interpret the neutron scattering data. Therefore according to fluctuation-dissipation theorem, the interpretation of neutron scattering experiment requires, only an understanding of the target itself. In this section following \cite{Lovesey2}, we briefly summarize the formulation of interaction of neutrons with matter. 

  The wave-length in $\AA$ and energy in $meV$ of neutrons are related by 
\be
  \lambda = 9.04 ~ E^{-1/2}
\ee
Low energy neutrons are often described as cold, thermal, hot, or epithermal. The typical energy scales and corresponding de Broglie wave-lengths are given below
\be
\begin{array}{|l|l|l|}
  \hline
                    & E/ \mbox{meV}   & \lambda / \AA\\
  \hline \hline
  \mbox{Cold}       & 1-10            & 9-3          \\
  \hline
  \mbox{Thermal}    & 10-100          & 3-0.9        \\
  \hline
  \mbox{Hot}        & 100-500         & 0.9-0.4      \\
  \hline
  \mbox{Epithermal} & 500-            & 0.4 -        \\
  \hline
\end{array}
\ee

 A simple golden rule for the scattering of neutrons gives the cross section proportional to dynamical form factor $S({\bq},\omega)$, which is itself proportional to $\mbox{Im}\chi({\bq},\omega)$. The dynamical susceptibility $\mbox{Im}\chi({\bq},\omega)$, measured by neutron scattering yields incisive information about the electron-electron interactions on precisely the relevant energy and length scales. It is widely considered as one of the most important probes of strongly correlated materials.

  Assume that the incident neutron gets scattered from initial plane-wave state $|{\bk}\rangle$ to $|{\bk'}\rangle$, whereby the target acquires a transition form initial state $|n\rangle$ to final state $|n'\rangle$. Within the Born approximation, the contribution form this process to total cross-section is given by \cite{Lovesey2}
\bearr
   \left(\frac{d^2\sigma}{d\Omega dE'}\right)_{n\to n'} = 
   \frac{k'}{k}\left|\langle{\bk'}n'|\hat{V}|{\bk}n\rangle \right|^2
   \delta(\hbar\omega-E_{n'n})
\eearr
where $E_{n'n}=E_{n'}-E_{n}$ is the energy change of target, and
\bearr
   \langle{\bk'}|\hat{V}|{\bk}\rangle &=&\frac{m_N}{2\pi\hbar^2}
   \int d{\br}~e^{i({\bk}-{\bk'}).{\br}}~V({\br})
\eearr
For none zero temperatures, there is a range of accessible initial states with probability $p_{n} \sim e^{-\beta E_{n}}$. The basic quantity which is measured, is the partial differential cross section
\be
   \frac{d^2\sigma}{d\Omega dE'}=\frac{k'}{k} \sum_{n} p_{n}
   \left|\langle{\bk'}n'|\hat{V}|{\bk}n\rangle \right|^2
   \delta(\hbar\omega-E_{n'n})\label{cross-section.eqn}
\ee
In the above expression $\hat{V}$ is the interaction potential of neutrons with the target. The neutrons may interact with ions either through nuclear forces of spin and orbital angular momentum dipole interactions, or, with electrons dipole. The case of relevant to our problem will be interaction with the spin dipole moment of $2p$ 'band electrons'.

  \subsection{Scattering by electrons}
If we forget about the orbital motion of electrons, the magnetic field of its spin dipole moment ${\bf\mu}_e=-2\mu_B\mathbf{s}$ becomes 
\be
   {\mathbf H}={\mathbf \nabla} \times \left({\bf\mu}_e \times 
   \frac{{\mathbf R}}{|{\mathbf R}|^3} \right)
\ee
so that the interaction potential between the electron and neutron becomes
\be
   -\gamma{\mathbf \mu}_N~ {\hat{\bf\sigma} .{\bf H}}=
   2\gamma{\mathbf\mu}_N \mu_B ~{\hat{\mathbf{\sigma}}} .{\mathbf{\nabla}} 
   \times \frac{{\mathbf s} \times {\mathbf R}}{|{\mathbf R}|^3}
\ee
where $\bf{\hat{\sigma}}$ denotes the Pauli spin operators. Plugging it into equation (\ref{cross-section.eqn}) and after some algebra \cite{Lovesey2} we find
\bearr
  &&\frac{d^2\sigma}{d\Omega dE'} = \left(\frac{m_N}{2\pi\hbar^2}\right)^2
  (2 \gamma \mu_N \mu_B)^2 (4\pi)^2 \frac{k'}{k} 
  \sum_{nn'\sigma\sigma'}p_{n}p_{\sigma}\nn\\
  &&\langle n\sigma|({\bf\hat{\sigma}}.{\bf\hat{Q}})^{\dagger}
  |n'\sigma'\rangle
  \langle n'\sigma'|{\bf\hat{\sigma}}.{\bf\hat{Q}}|n\sigma\rangle
  ~\delta(\hbar\omega-E_{n'n})\label{cross2.eqn}
\eearr
where the operator ${\bf\hat{Q}}$ is defined by
\be
   {\bf\hat{Q}}=\sum_{i} e^{i{\bq}.{\br}_i} \tilde{\bq}\times
   ({\bf\hat{s}}_i\times \tilde{\bq})
\ee
Here ${\bq}={\bk'}-{\bk}$ is the momentum transfer, and $\tilde{\bq}={\bq}/|{\bq}|$ is the unit vector along ${\bq}$. Finally putting every thing together, we have
\bearr
  \frac{d^2\sigma}{d\Omega dE'} &=& r_0^2 \frac{k'}{k} S({\bq},\omega)
  \label{dsigma.eqn}
\eearr
where $r_0=\gamma e^2/m_ec^2$ and the structure factor $S({\bq},\omega)$ is given by
\bearr
  && \sum_{\alpha\beta} f_{\alpha\beta}
  \sum_{nn'}p_{n}\langle n|\hat{Q}_{\alpha}^{\dagger} |n'\rangle
  \langle n'|\hat{Q}_{\beta}|n\rangle ~\delta(\hbar\omega-E_{n'n})
  \label{struct1.eqn}
\eearr
 The coefficients $f_{\alpha\beta}$ depend on whether we are using polarized or unpolarized neutron beam and are given by
\be
   f_{\alpha\beta}=\sum_{\sigma}p_{\sigma}
   \langle\sigma|\hat{\sigma}_{\alpha}\hat{\sigma}_{\beta}|\sigma\rangle
\ee
For experiments with unpolarized neutrons $f_{\alpha\beta}=\delta_{\alpha\beta}$ while for polarized neutrons, say with polarization $|\up\rangle$,
\be
f_{\alpha\beta}=\left(
\begin{array}{lll}
  1  & i  & 0\\
  -i & 1  & 0\\
  0  & 0  & 1
\end{array}
\right) =\left(
\begin{array}{lll}
  1  & 0  & 0\\
  0  & 1  & 0\\
  0  & 0  & 1
\end{array}
\right)+\left(
\begin{array}{lll}
  0  & i  & 0\\
  -i & 0  & 0\\
  0  & 0  & 0
\end{array}
\right)
\ee
where the second term really distinguishes polarized experiments form 
unpolarized ($\delta_{\alpha\beta}$) one. In polarized neutron scattering 
investigation of spin-1 excitations, one counts only the neutrons that after 
scattering, have flipped their spin. However, fortunately in case of 
graphite the spin collective mode exists up to electronic energies of the 
order of $eV$, where there are no phonon degrees of freedom (phonons inhabit 
in energy scale tens of $meV$) with comparable energies. So one can safely 
do the experiments with {\em unpolarized} neutrons.  With these assumption e
quation (\ref{struct1.eqn}) can be written in terms of the Fourier component 
of the spin density operator as
\bearr
  \sum_{\alpha\beta} 
  (\delta_{\alpha\beta}-\tilde{q}_{\alpha}\tilde{q}_{\beta})
  \frac{1}{2\pi\hbar}\int_{-\infty}^{\infty} dt~e^{-i\omega t}
  \langle S_{\bq\alpha}S_{\bq\beta}^{\dagger}(t)\rangle
  \label{struct2.eqn}
\eearr
where in the last step we have used the Fourier representation of the Dirac 
delta function in equation (\ref{struct1.eqn}) and the interaction picture 
representation of the spin operator, that is:
\be
  {\bf S}_{\bq} = \sum_{i} e^{ -i {\bq}.\br_i}  {\bf s}_i
\ee

  \subsection{Itinerant electrons}
For the band electrons, $S^{\dagger}_{\bq\alpha}$ is represented in Bloch 
basis as
\be
  S^{\dagger}_{{\bq}\alpha}=\frac{1}{2} \int d{\br} e^{i {\bq}.\br} 
  \sum_{{\bk\bk'}\lambda\lambda' ss'}
  \psi^{\lambda*}_{\bk}({\br})\psi^{\lambda'}_{\bk'}({\br})~
  c^{\lambda\dagger}_{{\bk} s} \hat{\sigma}^{\alpha}_{ss'} 
  c^{\lambda'}_{{\bk'}s'}\label{spinbloch1.eqn}
\ee
where $\lambda$ is the band index, $s=\up,\down$ is the spin projection and 
$\psi^{\lambda}_{\bk}$ represents the Bloch wave-function. Using equations 
(\ref{Bloch.eqn}) and (\ref{LCAO.eqn}), after some algebraic steps, the 
representation of the spin operator in Bloch basis of graphene becomes
\be
  S^{\dagger}_{{\bq}\alpha}=\frac{1}{2}
  \sum_{{\bk}\lambda\lambda' ss'} f^{\lambda\lambda'}_{{\bk},{\bq}} ~
  c^{\lambda\dagger}_{{\bk}+{\bq} s}
  \hat{\sigma}^{\alpha}_{ss'} c^{\lambda'}_{{\bk}s'}
  \label{spinbloch2.eqn}
\ee
where $(\lambda,\lambda'=\pm 1)$
\bearr
  f^{\lambda\lambda'}_{{\bk},{\bq}} &=& \frac{1}{2} F_0({\bq})\left[ 
  1+\lambda\lambda' e^{i\varphi({\bk}+{\bq})-i\varphi({\bk})}\right]\nn\\
  &&+\frac{\lambda'}{2}F_1({\bq})e^{-i\varphi({\bk})+i{\bk}.{\bf d} }\nn\\
  &&+\frac{\lambda}{2}F^*_1({-\bq})e^{i\varphi({\bk}+{\bq})-i({\bk}+{\bq}).
  {\bf d}}
  \label{form1.eqn}
\eearr
The form factors and phase factors are given by
\bearr
  e^{i\varphi({\bk})} &=& \frac{f({\bk})}{|f({\bk})|}\\
  F_0({\bq}) &=&\int d{\br}\left\vert\phi({\br})\right\vert^2e^{i{\bq}.{\br}}\\
  \label{F0Form.eqn}
  F_1({\bq}) &=&\int d{\br}~\phi^*({\br})e^{i{\bq}.{\br}}\phi({\br}-{\bf d})
\eearr
where $f({\bk})$ is defined in equation (\ref{fofk.eqn}). $\phi({\br})$ and 
$\phi({\br}-{\bf d})$ represent the atomic $p_z$ orbital at some carbon atom 
and its neighboring site, respectively. 

  Neglecting the form factor $F_1({\bq})$ which mixes the two neighboring 
orbitals, and using the symmetry between two sub-lattices of graphite, 
equation (\ref{spinbloch2.eqn}) can be written as
\be
  S^{\dagger}_{{\bq}\alpha}=
  F_0({\bq}) \sum_{{\bk}\lambda ss'}  ~ 
  c^{\lambda\dagger}_{{\bk}+{\bq} s} \frac{1}{2}
  \hat{\sigma}^{\alpha}_{ss'} c^{\lambda}_{{\bk}s'}
  \label{spinbloch3.eqn}
\ee
This is in fact the one-band result which is not surprising, because we have neglected the cross term between bands, and therefore the final result is same as one-band case, with of course summation over two separated bands.

  Using the fluctuation dissipation theorem \cite{Lovesey1,Lovesey2}, we have
\be
   \label{fluct-dissip-theorem.eqn}
   (1-e^{-\beta\omega})S(\bq,\omega)=
   -  \frac{1}{\pi} |F_0(\bq)|^2 \mbox{Im}\chi(\bq,\omega)
\ee
Hence, equation (\ref{dsigma.eqn}) can be written in the form of
\bearr
  \frac{d^2\sigma}{d\Omega dE'} &=&-\frac{r_0^2}{\pi(1-e^{-\beta\omega})}
  \frac{k'}{k} |F_0(\bq)|^2 \mbox{Im}\chi(\bq,\omega)
  \label{dsigmaFinal.eqn}
\eearr
 Therefor the peak structure of neutron scattering cross section is reflected in dissipative part of the susceptibility, with an overall atomic form factor, reflecting the band nature of electrons involved in the process. The Im$\chi(\bq,\omega)$ includes the effects of interaction among electrons. Treating the effect of interaction in RPA approximation, since the life-time effects are beyond RPA, Im$\chi^{RPA}$ has very isolated sharp peaks at resonance frequencies $\omega_s(\bq)$, that is
\be
   \mbox{Im}\chi^{RPA}(\bq,\omega) \approx Z(\bq)~
   \delta(\hbar\omega-\hbar\omega_s(\bq)) 
\ee
which defines a dimension-less quantity $Z(\bq)$. By Kramers-Kronig relation, the above equation is equivalent to 
\be
  \mbox{Re}\chi^{RPA}(\bq,\omega)\approx -\frac{1}{\pi\hbar} 
  \frac{Z(\bq)}{\omega-\omega_s(\bq)},~~~~~~\omega\approx\omega_s(\bq)
\ee
Within the RPA, one does not need to perform any fitting to obtain $Z(\bq)$. 
Upon using equation (\ref{RPA.eqn}), $Z(\bq)$ can be obtained in terms of 
{\em bare} susceptibility as
\be
   Z(\bq)=\frac{\pi\hbar}{U^2} \left[\frac{\partial}{\partial \omega}
   \mbox{Re}\chi^0(\bq,\omega)\right]_{\omega=\omega_s(\bq)}^{-1}
   \label{peak1.eqn}
\ee
The measured intensity at an angle corresponding to momentum transfer $\bq$, apart from overal factors is given by 
\be
   I(\bq)=|F_0(\bq)|^2 Z(\bq)
   \label{Iq.eqn}
\ee
  Now let us discuss the effect of atomic form factors, which directly affects the intensity of the neutron peaks. For $2p_z$ atomic orbital, we have
\be
  \phi({\br})\sim z~e^{-Zr/2a_0} \label{pzcarbon.eqn}
\ee
where $a_0$ is the Bohr radius and $Z$ is the {\em effective} nuclear charge, which for $2p$ orbitals in carbon is $\approx 1.56$. {\em Assuming that, there is no momentum transfer along the direction perpendicular to graphene planes,} the atomic form factor becomes
\be
  F_0({\bq})=\frac{1}{[1+\frac{q^2 a_0^2}{Z^2}]^3}\approx
	\frac{1}{[1+\frac{q^2 a^2}{(7.8)^2}]^3}
  \label{F0q.eqn}
\ee
  Note that $a_0\approx a/5$ where $a$ is the lattice parameter as depicted in figure \ref{lattice.fig}. For momentum transfers much away from the center of the BZ, the weight of the neutron scattering peak is going to be reduced by $|F_0(\bq)|^2$. 

  For high energy part of the collective mode branch, we will need high energy incident neutron beam. Considering the neutron kinematics, the lowest $q$-values which can be attained (at the lowest possible scattering angles of $\sim 2-3 $ degree) are relatively high, i.e. much higher than the $q-$vector of the first Brillouin zone. Hence, one needs to know the magnitude of the form factor at $q-$values from $\sim 5-12\AA ^{-1}$ \cite{Murani} which corresponds to $qa\sim 12-29.5$. The above estimate of the atomic form factor indicates that the intensity of peaks is essentially zero beyond the first BZ!. This makes the detection of the high energy parts a technically demanding task and the appropriate region of the BZ should be chosen carefully. Keeping this consideration in mind, after calculating the coefficient of the delta peaks of Im$\chi(\bq,\omega)$ in RPA approximation, we will suggest the best regions of the BZ (figure \ref{peak.fig}) to be probed by neutrons.

\section{the tight-binding Band structure}
\label{TightBinding.sec}
Within the tight-binding approximation, and neglecting the overlap of $2p_z$ atomic orbitals of neighboring carbon atoms, the band dispersion of graphene are given by \cite{Saitobook,Wallace} $\varepsilon^{c/v}_{\bk} = \pm t~\varepsilon_{\bk}$ where

\begin{figure}[ht]
   \begin{center}
   \includegraphics[scale=0.45]{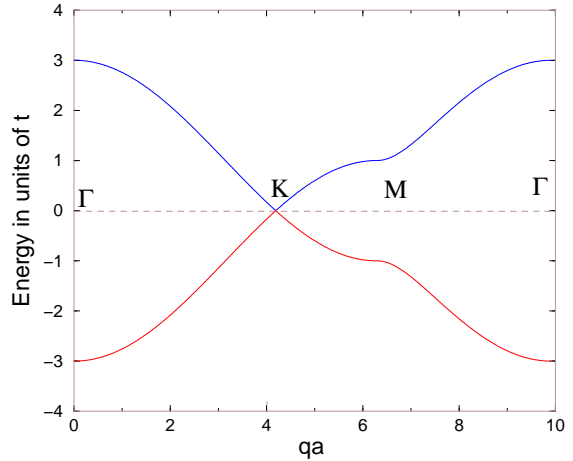}
   \vskip -0.1 cm
   \caption{$\pi-$bands of graphene along $\Gamma KM\Gamma$ loop, at tight-binding approximation. The $M-$point is a saddle point and gives rise to a van-Hov singularity in single-particle DOS. In this figure energy is in units of $t$ and the horizontal axis denotes $qa$, with $a$ defined in figure \ref{lattice.fig}}
   \label{vband.fig}
   \end{center}
\end{figure}
\begin{figure}[t]
\begin{center}
\includegraphics[scale=0.30]{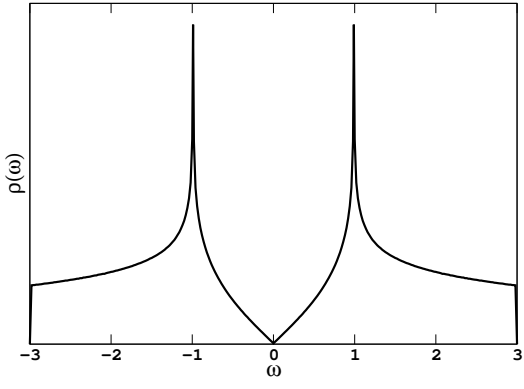}
\vspace{0.5 cm}
\caption{The DOS corresponding to tight-binding band equation(\ref{band.eqn}). The energies are in units of $t$. The linear behavior for small energies results form the cone-like features of the band structure near $K-$points and the saddle point at $M-$point is responsible for the cusp in DOS at $|\omega|=1$}
\label{dos.fig}
\end{center}
\end{figure}


\be
   \varepsilon_{\bk} =
   \sqrt{1+4\cos(\frac{\sqrt 3 k_x a}{2})\cos(\frac{k_y a}{2})+
   4\cos^2(\frac{k_y a}{2})} \label{band.eqn}
\ee
where $+$ and $-$ signs correspond to conduction and valence bands, respectively. Here $a$ is the length of translation vector in one of the sub-lattices of graphene: $a=\sqrt 3 \times$(C-C bond length) as in figure \ref{lattice.fig}. The above bands touch each other at the $K-$points (corners) of the BZ as depicted in figure \ref{vband.fig}. Figure \ref{dos.fig} shows the DOS of the  dispersion (\ref{band.eqn}). 

 For later reference, the normalized Bloch orbitals of graphene at tight-binding approximation become \cite{Wallace}
\bearr
  \psi_{\bk}^{\pm}({\bf r})
  &=&\frac{1}{\sqrt N}\sum_{{{\bf R}} \in A} {e^{i {\bf k} . {\bf R}}} 
  \phi_{\bf k,R}^{\pm}({\bf r})\label{Bloch.eqn}\\
  \phi_{\bf k,R}^{\pm}({\bf r}) &=&\frac{1}{\sqrt 2}\left[\phi({\bf r}-{\bf R})
  \pm \frac{\vert f(\bk )\vert}{f(\bk)}e^{i{\bf k} . {\bf d}} 
  \phi({\bf r}-{\bf R}-{\bf d})\right]\nn\\ 
  &&\label{LCAO.eqn}
\eearr
with
\be
   f({\bf k}) = e^{i\frac{k_x a}{\sqrt 3}}+2e^{-i\frac{k_x a}{2\sqrt 3}}
   \cos(\frac{k_y a}{2}) \label{fofk.eqn}
\ee
where ${\bf R}$ runs over one of the sub-lattices of graphene and ${\bf R}-{\bf d}$ refers to carbon atom in neighboring site which lives in the other sub-lattice. Here $\phi$ is normalized $p_z$ atomic orbital.

The above bands when linearized around $K-$points of the BZ become
\be
   \label{cone.eqn}
   \varepsilon_{\bk}=\left\{
   \begin{array}{ll}
  +\hbar v_F \vert \bk \vert &~~~~~~\mbox{conduction band}\\
  -\hbar v_F \vert \bk \vert &~~~~~~\mbox{valence band}
   \end{array}\right.
\ee
where $\hbar v_F = \frac{\sqrt 3}{2}t a$. The DOS corresponding to the this linear dispersion is given by 
\[
   \rho(\varepsilon)=\frac{1}{2\pi\hbar^2 v_F^2}|\varepsilon|
\]
So the $\rho(\varepsilon)\propto|\varepsilon|$ pseudo-gap arises form the linear features of the band structure near the $K-$points. It can be shown that, within DMFT, this cone-like feature remains robust against the increase in the on-site repulsion $U$ \cite{Diploma}. Hence the on-site interaction does not destroy the Dirac cone picture. It only renormalizes the Fermi velocity $v_F$. Figures \ref{vband.fig} and \ref{dos.fig} depict the $\pi$ bands of graphene and corresponding DOS.

\section{The RPA susceptibility}
\label{RPAchi.sec}
  In order to calculate the contributions to total RPA susceptibility form different channels, we need the (retarded) {\it bare susceptibility} which is given by
\be
   \chi^0(\bq,\omega)=\frac{1}{N}\sum_{\bk}\frac{f_{\bk+\bq}-f_{\bk}}
   {\hbar\omega-(\varepsilon^c_{\bk+\bq}-\varepsilon^v_{\bk})+i0^+}
   \label{starting.eqn}
\ee
where $N$ is the number of unit cells. At $T=0$ the conduction band is empty and we have $f_{\bk + \bq}=0$. At half filling which is the case for undoped graphene the valence band is completely filled and we have $f_{\bk}=1$. Therefore the bare susceptibility becomes (energies are in units of $t$, unless otherwise specified)
\bearr
   \chi^0(\bq,\omega) &=& \frac{1}{N}\sum_{\bk}\frac{-1}
   {\hbar\omega-(\varepsilon^c_{\bk+\bq}-\varepsilon^v_{\bk})+i0^+}\nn\\
   &=& \frac{1}{N}\sum_{\bk}\frac{1}
   {(\varepsilon_{\bk+\bq}+\varepsilon_{\bk})-\hbar\omega-i0^+}
   \label{chi.eqn}
\eearr

The imaginary part of $\chi^0$ upon using the formula 
\[
\frac{1}{x-i0^+}={\cal P}\frac{1}{x}+i\pi\delta(x)
\]
can be written as
\bearr
   \label{imchi.eqn}
   \mbox{Im}\chi^0(\bq,\omega) &=& \frac{\pi}{N}\sum_{\bk}
   \delta\left[\hbar\omega-(\varepsilon^c_{\bk+\bq}-\varepsilon^v_{\bk})
   \right]\nn\\
   &=&\frac{A}{4\pi}\int_{BZ}d^2\bk~
   \delta\left[\hbar\omega-(\varepsilon_{\bk+\bq}+\varepsilon_{\bk})\right]
\eearr
where $A=\frac{\sqrt 3 a^2}{2}$ is the area of hexagonal unit cell. Using Kramers-Kronig relation, the real part of $\chi^0$ can be obtained from the imaginary part 
\be
   \mbox{Re}\chi^0(\bq,\omega)=\frac{1}{\pi}{\cal P}\int_{-\infty}^{+\infty}
   d\omega'\frac{\mbox{Im}\chi^0(\bq,\omega')}{\omega'-\omega}
   \label{Kramers.eqn}
\ee

\begin{figure}[t]
\begin{center}
\includegraphics[scale=0.45]{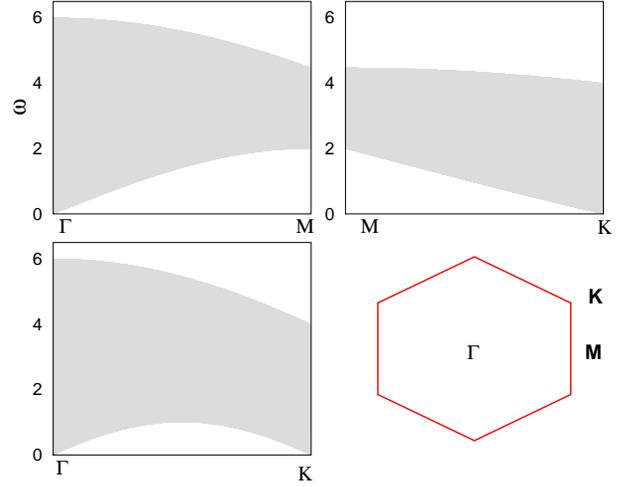}
\vspace{0.5 cm}
\caption{Particle-hole continuum of graphite. Horizontal axis is center of mass momentum of particle-hole pair and vertical axis is the particle-hole energy. Gray region is the particle-hole continuum, for the spectrum given by equation (\ref{band.eqn}). The energies are in units of $t$. This is obtained for a fixed ${\bq}$, by finding the maxima and  minima of the particle-hole energy $\varepsilon_{\bq+\bk}+\varepsilon_{\bk}$ for random walker $\bk$ walking in Brillouin zone.}
\label{ph_continuum.fig}
\end{center}
\end{figure}
The particle-hole continuum is a region in $\omega -\bq$ space in which integrand of equation (\ref{imchi.eqn}) is none-zero. Out of this region the denominator of summand which gives $\chi^0$ is none-zero and the integral becomes well-behaved and easy to be performed numerically.

  The contribution form the triplet particle-hole channel to the RPA susceptibility (particle-hole ladder summation) is given by the equation \cite{FLEX}
\be
  \chi(\bq,\omega)=\frac{\chi^0(\bq,\omega)}{1-v(\bq)\chi^0(\bq,\omega)}
  \label{rpachi.eqn}
\ee
where $v(\bq)$ is the Fourier component of potential. The short range part of the interaction is sufficient to account for the spin phenomena. Hence we use the standard Hubbard model for this tight-binding system,
\be
   H=-t\sum_{<i,j>,\sigma}
   (\cdag_{i\sigma}c_{j\sigma}+\cdag_{j\sigma}c_{i\sigma})
   +U\sum_{i}n_{i\up}n_{i\down}
\ee
with repulsive interaction between the particles $U>0$. For this interaction $v(\bq)=U$, and the contribution form triplet particle-hole channel to the RPA susceptibility becomes
\be
  \chi(\bq,\omega)=\frac{\chi^0(\bq,\omega)}{1-U\chi^0(\bq,\omega)}
  \label{RPA.eqn}
\ee
  The collective mode can exist if the above RPA susceptibility diverges. The roots of the denominator are the solutions to equations
\be
   \label{SZS.eqn}
   \mbox{Im}\chi^0(\bq,\omega)=0,~~~~~~
   \mbox{Re}\chi^0(\bq,\omega)=\frac{1}{U}
\ee
 One can see from equation (\ref{chi.eqn}) that out of particle-hole continuum where $\hbar \omega \neq \varepsilon_{\bk+\bq}+\varepsilon_{\bk}$, we have Im$\chi^0=0$, i.e. $\chi^0$ is purely real. Moreover below the particle-hole continuum $\hbar \omega < \varepsilon_{\bk+\bq}+\varepsilon_{\bk}$ and thus from equation (\ref{chi.eqn}) one can see that $\mbox{Re}\chi^0=\chi^0 > 0$. This shows that {\em below the particle-hole continuum}, there could be a solution to the collective mode equations in triplet particle-hole channel. 

  Hence opening of a window below the particle-hole continuum provides the unique opportunity for the existence of a solution to equation (\ref{SZS.eqn}). The only remaining part is to see if this solution is out of the particle-hole continuum or inside it. To this end we need to solve equation (\ref{SZS.eqn}) numerically. Our strategy is to fix $\bq$ and then for a given $U$ look for a value of omega which fulfills the collective mode criterion, equation (\ref{SZS.eqn}). Figure \ref{collective.fig} shows the numerical solution of the collective mode equation. 

  Before going into the numerics, it is instructive to work out a simple model which is relevant to the low-energy physics of the problem. The asymptotic behavior of the collective mode dispersion near $\Gamma$ and $K-$points can be understood in terms of a very simple model. The clue is that near the $K-$points of the BZ, the spectrum is (Dirac) cone-like and this linearity holds true up to $\sim 0.5 t$. So as a model it seems plausible  to start with the simplified model of Dirac fermions, living in a circular BZ and subject to Hubbard interaction. It turns out that this problem will teach us lots about the nature of spin collective mode.

\section{simplified model of cone-like bands in circular BZ}
\label{CONE.sec}  
  In this section we use the linearized spectrum of equation 
(\ref{cone.eqn}) as a model to evaluate $\chi^0$ analytically. 
We assume that for this model BZ is a circle of radius $k_c$ 
which for the low-energy regime can be thought of as infinity. 
We assume that there is 
a {\em single cone} \footnote{A covariant formulation in terms of 
Dirac spinors, should take care of $2$ cones, $2$ chiralities, 
and $2$ spin degrees of freedom, which requires a $2^3=8$ component 
formalism.} at the center of the circular BZ. Note that in real 
graphite there are {\em two} such a cones. Such a simplified model 
seems to capture the essential low-energy physics of band 
structure (\ref{band.eqn}). The linear spectrum of equation 
(\ref{cone.eqn}) holds true over a very large range of energies 
(up to $\epsilon\sim 0.5 t$) which can be seen qualitatively 
from the density of states (figure \ref{dos.fig}). Such a model 
will be relevant to
\begin{itemize}
\item Wave vectors near $K-$point, of course with modified 
parameters such as Fermi velocity $v_F$, etc.
  \item Wave vectors near $\Gamma-$point. The analytical knowledge gained about the small$-q$ behavior of the dispersion of spin collective mode will enable us to see the qualitative difference between the case of a isolated graphene sheet and a real graphite in which the long-range (small$-q$) tail of the interaction becomes essential. We will include these effect analytically and will see that at $\Gamma-$point which is susceptible for instability, the collective mode still survives. 
\end{itemize}
Moreover, it will teach us about a mechanism by which the system exhibits 1D characters. It turns out that a very same mechanism is responsible for the existence of the spin collective mode in almost large fraction of BZ area.

  Starting with equation (\ref{imchi.eqn})
\bearr
   \mbox{Im}\chi^0(\bq,\omega) &=& \frac{A}{4\pi}\int_{|\bk|<k_c}
   \delta\left[\hbar\omega-\hbar v_F(|\bk+\bq|+|\bk|)\right] \nn \\
   &=&\frac{A}{4\pi\hbar v_F}\int 
   \frac{ds}{|\nabla_{k} f_{\bq}(\bk)|_{f_{\bq}(\bk)=z}}
\eearr
where
\[
   f_{\bq}(\bk)=|\bk-\bq|+|\bk|,~~~~~~z=\frac{\omega}{v_F}
\]
and $ds$ is the length element on $f_{\bq}(\bk)=z$. The equation $f_{\bq}(\bk)=|\bk-\bq|+|\bk|=z$ defines and ellipse with principal axis equal to $\frac{q}{2}$ and half conical distance equal to $\frac{z}{2}=\frac{\omega}{2 v_F}$. Therefor Im$\chi^0$ is none zero if and only if
\be
   z>q \Rightarrow \omega > v_F q~~~~~~\mbox{particle-hole continuum}
\ee
This result should be compared with figure \ref{ph_continuum.fig} around $K-$point and $\Gamma -$point.

  Assuming that $\phi$ is the angle between $\bq$ and $\bk$ and $k=|\bk|$, the equation of ellipse becomes
\be
   k=\frac{z^2-q^2}{2(z-q\cos\phi)}
\ee
which gives 
\bearr
   dk &=& -\frac{q(z^2-q^2) \sin \phi}{2(z-q\cos\phi)^2}d\phi \nn\\
   ds &=& \sqrt{k^2d\phi^2+dk^2}\nn\\
   &=& \frac{z^2-q^2}{2(z-q\cos\phi)^2}\sqrt{z^2+q^2-2zq \cos\phi}\nn\\
   |\nabla_k f_{\bq}(\bk)| &=& 
   \frac{2(z-q\cos\phi)}{\sqrt{z^2+q^2-2zq \cos\phi}}
\eearr
Putting every thing together we obtain
\bearr
   && \mbox{Im}\chi^0(\bq,\omega) = \frac{z^2-q^2}{16\pi\hbar v_F}
   \int \frac{z^2+q^2-2zq \cos\phi}{(z-q\cos\phi)^3}~d\phi \nn\\
   &=& \frac{1}{16\hbar v_F}\frac{2z^2-q^2}{\sqrt{z^2-q^2}}\nn\\
   &=& \frac{\sqrt 3 a^2}{16\hbar v_F^2}
   \frac{\omega^2-\frac{1}{2}v_F^2 q^2}{\sqrt{\omega^2-v_F^2 q^2}},
   ~~ q v_F < \omega < (2k_c+q) v_F
\eearr
where in last step we have used $A=\frac{\sqrt 3}{2}a^2$ to restore the lattice parameter $a$. The real part is obtained using the Kramers and Kronig relation, equation(\ref{Kramers.eqn}), along with change of variables $\omega'=q v_F \coth\eta,~\coth\eta_0=1+\frac{2k_c}{q}$. To leading order in $\frac{2k_c}{q}$ we obtain
\bearr
   &&\hbox{Re}\chi^0(\bq,\omega) = \frac{\sqrt 3 a^2}{32\hbar\pi v_F^2}
   \int_{qv_F}^{2k_c v_F} \frac{d\omega'}{\omega'-\omega}
   \frac{2\omega^2-v_F^2 q^2}{\sqrt{\omega^2-v_F^2 q^2}}\nn\\
   &=&-\frac{\sqrt 3 a^2 q^2}{32\pi\hbar}\int_{\eta_0}^{\infty}d\eta
   \frac{2\coth^2\eta-1}{\sinh\eta(\omega-q v_F\coth\eta)}\nn\\
   &=& -\frac{\sqrt 3 a^2}{16\pi\hbar v_F^2} 
   [\omega\log(\frac{4k_c}{q})-qv_F (\frac{2k_c}{q}+1)\nn\\
   &&+\frac{q^2v_F^2-2\omega^2}{\sqrt{q^2v_F^2-\omega^2}}
   (\arctan\sqrt{\frac{qv_F-\omega}{qv_F+\omega}}-\frac{\pi}{2})]
   +{\cal O}(\frac{q}{k_c})\nn\\
   &=&\frac{\sqrt 3 k_c a^2}{8\pi\hbar v_F} +
   \frac{\sqrt 3 a^2}{16\pi\hbar v_F^2}
   \frac{2\omega^2-q^2v_F^2}{\sqrt{q^2v_F^2-\omega^2}}
   \arctan\sqrt{\frac{qv_F+\omega}{qv_F-\omega}}\nn\\
   &&+ {\cal O}(\log\frac{k_c}{q}),~~~~~~q v_F > \omega
   \label{rechi.eqn}
\eearr
  Now going back to equations (\ref{SZS.eqn}) we see that in region $q v_F > \omega$ where the above formula for Re$\chi^0(\bq,\omega)$ is valid, the imaginary part is identically zero. So the dispersion of collective mode is the solution to equation
\be
   \frac{1}{U}=\frac{\sqrt 3 k_c a^2}{8\pi\hbar v_F} 
   +\frac{\sqrt 3 a^2}{16\pi\hbar v_F^2}
   \frac{2\omega^2-q^2v_F^2}{\sqrt{q^2v_F^2-\omega^2}}
   \arctan\sqrt{\frac{qv_F+\omega}{qv_F-\omega}}
\ee
Near the particle-hole boundary the RHS of the above equation can be expanded to obtain (note that $\hbar v_F=ta\sqrt{3}/2 $)
\be
   \frac{1}{U}=\frac{ k_c a}{4\pi t} 
   + \frac{\sqrt 3 a^2 q^{3/2}}{32\hbar 
   \sqrt{2 v_F}\sqrt{qv_F -\omega}} + {\cal O}(qv_F-\omega)^{1/2}
   \label{rechiAssiptotic.eqn}
\ee
The above equation has a solution, passing through the origin for {\em any} value of $U < U_1=\frac{4\pi t}{k_c a}$. To get some idea about the value of $U_1$, let us choose $k_c$ in such a way that the area of our circular BZ is equal to the area of hexagonal BZ of the original problem, that is
\[
   \pi k_c^2=\frac{8\pi^2}{\sqrt 3~a^2}\Rightarrow 
   k_c a =2 \left( \frac{2\pi}{\sqrt 3} \right)^{1/2}\approx 3.81
\]
which gives $U_1=\left(2\pi\sqrt 3 \right)^{1/2}t\approx 3.30t$. Hence to leading order in $qa$ the dispersion of the spin-1 collective mode near the $\Gamma-$point becomes 
\be
   \label{dispersion.eqn}
   \omega_s(\bq) = v_F q(1-\alpha^2 a^2 q^2)=v_F q -\omega_B(q),
\ee
as $\omega \to v_F q\to 0$ with
\[
   \alpha=\frac{1} {8\sqrt 6}\frac{U U_1}{t(U_1-U)}
\]
 Note that equation (\ref{dispersion.eqn}) makes sense only if $\left|\alpha^2 a^2 q^2 \right| \ll 1 $. In particular the smallness of $\alpha$ implies that $U$ must be below $U_1$ and far enough from it. In fact, numerical calculations for the full band structure shows that for $U>U_c \approx 2.23~t$, an instability emerges around the $\Gamma-$point (see figure \ref{collective.fig}). Since the linear model is relevant to $\Gamma-$point (and also the $K-$point, but with {\em renormalized $v_F$}), the requirement of $U<U_1$, warns us of an instability. However, since the real graphite is stable, the normalized value of interaction must be less than $2.23~t$ and we are not worried about it.

Here $\hbar\omega_B$ is the {\em binding} energy of particle-hole pairs with center of mass momentum $\bq$. The repulsive interaction $U$ among particles becomes attractive for particle-hole pairs in triplet spin state and binds them together to form a bound state with binding energy $\hbar\omega_B$. To see why there exists a bound state for arbitrarily small attraction $U$, note that, right above the particle-hole continuum and very close to the continuum boundary, Im$\chi^0$ can be written as
\be
   \label{imchiAssiptotic.eqn}
   \mbox{Im}\chi^0(\bq,\omega)=\frac{\sqrt 3 a^2}{32\hbar \sqrt{2 v_F}}
   \frac{q^{3/2}}{\sqrt{\omega-q v_F }}
\ee
with a square root divergence at the lower edge of the particle-hole continuum
in $\omega -\bq$ space. This expression has the same form as the one 
dimensional density of states (with energy measured from $\hbar q v_F $).
Note that, in fact, Im$\chi^0(\bq,\omega)=\pi \rho_{\bq}(\omega)$, where 
$\rho_{\bq}(\omega)$ is the free particle-hole DOS for a fixed center of mass 
momentum $q$. {\em That is , the particle-hole pairs have a phase space for scattering which 
is effectively one dimensional}. Thus we have a bound state of particle-hole 
pairs in spin-triplet channel for arbitrarily small interaction $U$. However, 
we also have a pre factor $q^{3/2}$ which scales the density of states. This 
together with the square root divergence of the density of states at the 
bottom of the particle-hole continuum gives us a bound state for every $\bq$ 
as $q\to 0$, with the binding energy vanishing as $\sim q^3$ as shown above. 
Equation (\ref{imchiAssiptotic.eqn}) shows that entire mechanisms responsible 
for the formation of spin-1 zero sound (SZS) near the $\Gamma$ and 
$K-$points, is due to a region near the bottom of the particle-hole continuum.
In fact for center of mass momenta close to $\Gamma, K-$points, there will be 
a one dimensional manifold on which the particle-hole energy has its minimum 
and this {\em one-dimensional manifold of minima} is responsible for square root 
divergence in particle-hole DOS. The $M-$point has a similar property. 
These explains why for center of mass momentum corresponding to $M-$point 
too, the spin-1 collective mode is there for arbitrarily small interaction $U$.

\subsection{neutron resonance peaks for the cone model}
\label{NeutronCONE.sec}
 
  Plugging equations (\ref{rechi.eqn}) and (\ref{dispersion.eqn}) into the above equation, it is straightforward to show that near the $\Gamma$ and $K-$point the peak intensity behaves like
\bearr
   \label{peak2.eqn}
   Z(\bq) &=& \frac{3\pi^4}{4}\frac{U t^2}
   {\left(U_1-U \right)^3}\left(\frac{q}{k_c}\right)^3\times\nn\\
   &&\hspace{-0.9 cm}\left\{1-15\pi^2\frac{ U^2}
   {\left( U_1-U\right)^2} \left(\frac{q}{k_c}\right)^2 
   +{\cal O}\left(\frac{q}{k_c}\right)^3\right\}
\eearr
This equation shows that for a graphene layer the resonance peak residue, $Z(\bq)$ vanishes as $\sim q^3$, where $\bq$ is measured from $\Gamma$ or $K-$points of the BZ. The atomic form factor is essentially constant within the first BZ and therefore the cross section for momentum transfers near $\Gamma$ and $K$ are very small. Vanishingly small $Z(\bq)$ according to appendix A, means spatially large wave-function which can not be excited by neutrons with a typical de-Broglie wave-length of $\sim 1\AA$.

   Due to the presence of the other graphene layer and screening arising from the interlayer hopping among the layers, the power law $Z(\bq)\sim q^3$ may change in real 3D graphite. In next subsection, we will discuss the effect of semi-metallic screening, which amounts to taking into account the long range tail of the interaction among the electrons. 

\subsection{effect of long-ranged part of the interaction}
Having established the existence of a gapless spin-1 collective mode branch within Hubbard model and the RPA approximation, we will discuss whether the semi-metallic screened interaction of 3D stacked layers will affect our result. In tight binding situation like ours, the spin physics is mostly captured by the short range part of the repulsion among the electrons. So we do not expect a drastic change in the behavior of spin collective mode, if we correct the interaction, by including long range tails, i.e. modifications near ${\bq}=0$.  Below, we will show this explicitly and will discuss similar situations using an asymptotic analysis. 

  Interaction, including interlayer scattering between layers separated by distance $d$ is given by\cite{paco}
\bearr
   \tilde{v}(\omega,q) = \frac{2\pi e^2}{\epsilon_0 q}
   \frac{\sinh (q d)}
   {\sqrt{[\cosh(qd) + {\frac{2\pi e^2}{\epsilon_0 q} } 
   \sinh(qd)\chi_0(\omega,q)   ]^2 - 1 }}
   \label{interlayer.eqn}
   \nn
\eearr 
Starting from equation (\ref{starting.eqn}), one can see that the value of Re$\chi^0(\omega=0, \bq=0)$ is finite. Hence we have the leading behavior $\tilde{v}(0,q)\sim q^{-1/2}$. This behavior is something intermediate between the metallic screening, i.e., $\tilde{v}(q)\sim (q^2+k_{scr}^2)^{-1}$ and insulator behavior, i.e., $\tilde{v}(q)\sim q^{-1}$. Our argument below will assume the power low $\tilde{v}(q)\sim q^{-\nu} $ for long range behavior of screened interaction.

  The essential feature of the cone model which lead to a {\em gapless bound state} is the square root divergence in DOS of free particle-hole pairs. So let us encapsulate the essential features into the following form
\be
   \mbox{Im}\chi^0(\bq,\omega')\sim \frac{q^\eta}{\sqrt{\omega'-q}}
   \label{toyim.eqn}
\ee
which accommodates key features of the problem, namely, square root divergence and overall $q^\eta$ dependence. This enables us to study the interplay between these two features and their effect in modification of graphene solution. Assuming the collective mode dispersion to be 
\be
   \omega_s(\bq) = q - c q^\beta
   \label{dispersion2.eqn}
\ee
Using the Kramers-Kronig relation, the collective mode equation becomes
\bearr
   q^\nu&\sim&\frac{1}{\tilde{v}(q)} = \mbox{Re}\chi^0(\bq,\omega_s(\bq))\sim\int 
   \frac{d\omega'}{\omega'-\omega_s(\bq)}\frac{q^\eta}{\sqrt{\omega'-q}}\nn\\
   &=& q^\eta \int \frac{d\sqrt{\omega'-q}} {\omega'-\omega_s(\bq)}
   = q^\eta \int \frac{du}{u^2+q-\omega_s(\bq)}\nn\\
   &=& q^\eta \int \frac{du}{u^2+c q^\beta}\sim q^{\eta-\beta/2}\nn
\eearr
which gives
\be
   \beta=2(\eta-\nu)
   \label{exponent.eqn}
\ee
This equation is important in that, it determines the qualitative difference between the dispersion of the collective mode in graphene and graphite. In our case $\eta=3/2$. The graphene is characterized by $\nu=0$, or equivalently $\beta=3$ which is nothing, but equation (\ref{dispersion.eqn}). Once we add the other layers of graphene, i.e., $\nu=1/2$, we obtain $\beta=2$. This means that, at a given wave-vector, the effect of semi-metallic screening is to increase the binding energy. Another interesting observation is that an insulator type of screening ($\nu=1$) manages a collective mode with strictly linear dispersion but with slope $1-c$. So if $c$ is positive, the collective mode is below the particle-hole continuum and survives. However, if $c$ is negative, it enters the particle-hole continuum and decays into particle-hole pairs.  

  The case of {\em strictly one-dimensional} problem corresponds to $\eta=0$. So a none zero $\eta$ is a peculiar feature of our particular problem which should be taken into account in putting 1D-like features on formal basis in order to make it appropriate for bosonization. It is seen from equation (\ref{exponent.eqn}) that for the case of a real 1D problem $\beta=-2\nu$. Since in semi-metallic and insulator case $\nu > 0$, the exponent $\beta$ becomes negative and hence the expansion (\ref{dispersion2.eqn}) makes no sense. Therefor a bare 1D DOS (square root divergence) is not sufficient to bound electron-hole pairs together to form spin-1 zero sound (SZS). To conclude, {\em both square root and $q^{3/2}$ pre-factor are essential to manage a gapless spin-1 collective mode in graphite.}

Now let us use a similar asymptotic analysis to see how the leading behavior of $Z(\bq)$ gets modified by going from graphene to graphite. Differentiating Kramers-Kronig relation, equation (\ref{Kramers.eqn}) with respect to $\omega$ and using (\ref{toyim.eqn}) gives
\bearr
   &&\left.\frac{\partial}{\partial \omega}\chi^0(\bq,\omega)
   \right\vert_{\omega=\omega_s(\bq)} \sim q^\eta\int 
   \frac{d\sqrt{\omega'-q}}{(\omega'-\omega_s(\bq))^2}\nn\\
   &\sim& q^\eta \int \frac{du}{(u^2+q-\omega_s(\bq))^2}
   = q^\eta \int \frac{du}{(u^2+c q^\beta)^2}\sim q^{\eta-3\beta/2}\nn
\eearr
Using equations (\ref{peak1.eqn}) and (\ref{exponent.eqn}) we obtain
\be
   Z(\bq)\sim q^{2\eta-3\nu}
\ee
The case of graphene corresponds to $\eta=3/2$ and $\nu=0$ which gives rise to $q^3$ dependence in agreement with equation (\ref{peak2.eqn}). This equation says that in real graphite ($\nu=1/2$), $Z({\bq})\sim q^{3/2}$, i.e. the effect of semi-metallic screening is to strengthen the intensity of resonance peak at a given wave-vector. 

The above analysis reveals that, following properties are necessary to conspire to manage a spin-1 collective mode for the cone model of graphite:
\begin{enumerate}
   \item Dirac cone spectrum, i.e. a pseudo-gap of the form $g(\epsilon)=\left|\epsilon\right|$ to allow for a window below the particle-hole continuum {\em and} to produce a semi-metallic screening $\tilde{v}(q)\sim q^{-1/2}$ for 3D stack of graphite. 
   \item Square root divergence of the DOS of free particle-holes at the edge of particle-hole continuum, accompanied by $q^{3/2}$ measure which comes form two dimensionality of the original problem.
\end{enumerate}

\section{Numerical calculation of the weight of neutron scattering peaks}
\label{NumericBZ}

\begin{figure}[ht]
\begin{center}
\includegraphics[scale=0.45]{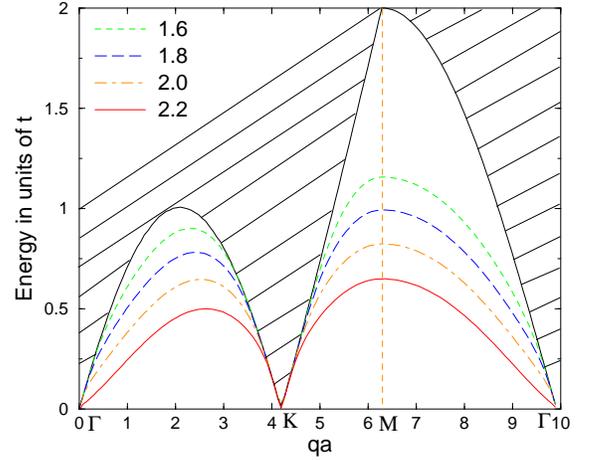}
\vspace{0.3 cm}
\caption{The spin-1 collective mode for different values of $U$. Note the emergence of instability at $\Gamma-$point at $U\approx 2.2$. Also note the asymptotic behavior of the dispersion of the spin-1 collective mode near $\Gamma$ and $K-$points. According to analysis of the single cone model, the collective mode dispersion near these points asymptotically approaches the continuum boundary. At $M-$point the resonance energy is at $\sim 0.83\times t\approx 2.0~eV$ and is well isolated from the boundary of p-h continuum. Lower energy regions are towards $\Gamma$ and $K$-points. Towards $\Gamma$-point binding energies are larger and better for neutron scattering as discussed in the text.}
\label{collective.fig}
\end{center}
\end{figure}

  So far our cone model which was relevant to small-$q$ behavior of the collective mode was especially suitable to address the physics near the $\Gamma-$point (and also $K-$point, with renormalized parameters), e.g. the question of modifications arising form inclusion of the long-range tail of interaction which arises form the stacks of 3D graphite. But for the neutron scattering experiments, we need to repeat the calculations in entire BZ. To find the resonance frequencies $\omega_s({\bq})$ in entire BZ, we have to numerically solve for equation (\ref{SZS.eqn}). In order to do so we have to perform the summation (\ref{chi.eqn}) or equivalently

\be
   \mbox{Re}\chi^0(\bq,\omega)=\frac{\sqrt 3 ~a^2}{8\pi^2}
   \int_{BZ} \frac{d^2\bk}
   {(\varepsilon_{\bk+\bq}+\varepsilon_{\bk})-\hbar\omega}
   \label{rchi.eqn}
\ee
where $A=\frac{\sqrt 3}{2}a^2$ is the area of units cell. We know form our analytical solutions that, below a certain value of $U$, the solution to collective mode equation (\ref{SZS.eqn}) lies below the particle-hole continuum and asymptotically approaches the lower boundary of continuum at $K$ and $\Gamma-$point, Hence the denominator of the integrand in the above equation in region in which we must look for the solutions $\omega_s(\bq)$ is none zero and with full band dispersion of equation (\ref{band.eqn}), the integral can be done numerically, without any problem. However, numerically it becomes more and more difficult to get the square root divergences as we go closer and closer to $K-$point. This is because of the pre-factor $q^{3/2}$ in equation (\ref{imchiAssiptotic.eqn}).

Figure \ref{collective.fig} shows the numerical evaluation of the dispersion of spin-1 collective mode for different values of $U$. As can be seen in figure \ref{collective.fig}, near $K-$point the collective mode pole is very close to the continuum boundary. One can fit a dispersion of type (\ref{dispersion.eqn}) near $\Gamma$ and $K-$points, but the coefficients will be different from that given by equation (\ref{dispersion.eqn}). Since near the $K-$point the binding energies are very small, the particle-hole pairs are bound loosely and the spatial extent of their wave-function is large which makes it difficult to excite them by neutrons with typical wave-length $\lambda\sim\AA$. {\em Therefore we suggest the neutron scattering experiments, NOT to focus around the $K-$points}. 
For the same reason the region {\em very} close to $\Gamma -$point should 
be avoided.

\begin{figure}[ht]
\begin{center}
\vspace{0.8 cm}
\includegraphics[scale=0.45]{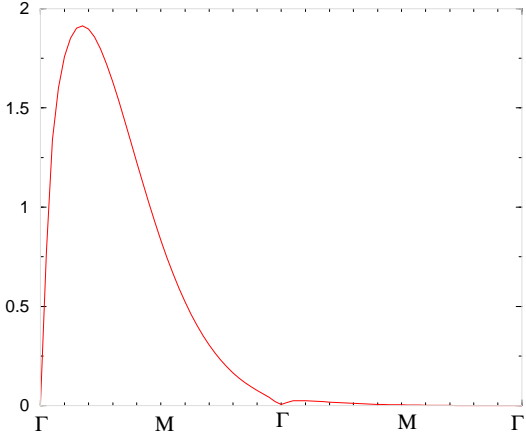}
\caption{Plot of $I(\bq)=|F_0(\bq)|^2 Z(\bq)$ as a function of momentum transfer along $\Gamma M$ direction. The $\Gamma\Gamma$ distance is $\approx 3\AA^{-1}$. Because of the effective nuclear charge $Z=1.56$ for $2p_z$ orbitals, the form factor decays very quickly, so that beyond the first BZ it is almost zero. This figure suggests that the M point is not suitable for neutron scattering. Since momentum transfers in $2^{nd}-4^{th}$ BZ ($5-12\AA^{-1}$) are involved at which peak intensity is almost zero (see the discussion following equation \ref{F0q.eqn}).}
\label{peak.fig}
\end{center}
\end{figure}

\begin{figure}[ht]
\begin{center}
\includegraphics[scale=0.75]{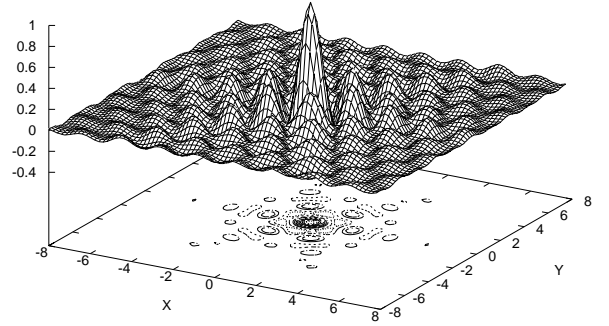}
\vspace{0.0 cm}
\caption{The approximate bound state wave-function within the RPA approximation. The center of mass momentum ${\bq}$ is at $0.2~\Gamma M\approx 0.3\AA^{-1}$. Unit of length is the lattice parameter $a\approx 2.46 \AA$. Contours corresponding to $|\psi_{\bq}({\br})|>0.1$ have been plotted in the base. Note a very soft anisotropic pattern of contours. The normalization of the wave-function is such that $\psi_{\bq}(0)=1$. This figure explicitly demonstrates why at low energies we have a {\em collective} mode. We expect $\sim 0.5~eV$ neutrons to be able to excite such a bound state.}
\label{waveGM0.2.fig}
\end{center}
\end{figure}

   Note that in figure \ref{collective.fig}, at $M-$point, there is a collective mode for any value of $U$. In fact at this point, the DOS of free particle-hole pairs diverges as $\sim 1/\sqrt{\omega-2t}$, where $2t$ corresponds to bottom of the particle-hole continuum at $M$. Bindings at $M-$point and its neighborhood, especially towards $\Gamma$, and also the point mid-way between $\Gamma K$ and region towards $\Gamma$ are strong enough to give rise to a particle-hole pair, with small enough wave functions that can be excited by thermal hot and epithermal neutrons. At $M-$point, the resonance energy is $\sim 2~eV$, but since at these energies momentum transfers beyond the first BZ are involved, the effect of atomic form factor reduces the intensity of the delta peak almost to zero. This makes neutron scattering at $M$-point a very difficult task. { Therefore despite the large binding energy, this point is NOT a good candidate to focus the neutron scattering experiments}. We will give a qualitative picture of wave-function in the following. 

Note the emergence of an instability at $\Gamma-$point near $U_c\approx 2.2t$. Indeed at the level of a Hartree-Fock mean field one finds a transition to FM phase at 
\[
   U_c=\frac{1}{\chi^0(\bq=0,\omega=0)}\approx 2.231t
\]
However, real graphite for which $U\sim 4t-5t$ is stable. So the value of $U_c=2.23 t$ should be an artifact of RPA approximation, and one expects by going beyond the RPA, to push $U_c$ above the $2.23 t$ or equivalently to obtain an screened value of $U$ below $2.23 t$ \cite{BJflex}. Assuming that the renormalized value of $U$ is less than $2.23 t$, we will do the rest of calculations for $U=2 t\sim 5~eV$.

  Once we find the location of resonance frequencies $\omega_s(\bq)$, the next step is to calculate $Z(\bq)$. The trick is to use the formula (\ref{peak1.eqn}), but with $\left[\frac{\partial}{\partial \omega} \mbox{Re}\chi^0(\bq,\omega)\right]_{\omega=\omega_s(\bq)}$ given by {\em direct differentiation of equation (\ref{rchi.eqn})}, that is
\bearr
   Z(\bq)&=&\frac{\pi}{U^2}\left(\sum_{\bk}\frac{1}
   {[\varepsilon_{\bk+\bq}+\varepsilon_{\bk}-\hbar\omega_s(\bq)]^2} 
   \right)^{-1} \label{zsum.eqn}\\
   &=&\frac{8\pi^3}{\sqrt 3~U^2 a^2}\left(\int \frac{d^2\bk}
   {[\varepsilon_{\bk+\bq}+\varepsilon_{\bk}-\hbar\omega_s(\bq)]^2}
   \right)^{-1} \label{zint.eqn}
\eearr
Then using equations (\ref{Iq.eqn}) and (\ref{F0q.eqn}) the weight of neutron scattering peaks $I({\bq})$, can be calculated, which is depicted in figure \ref{peak.fig}. The maximum value of the  $I({\bq})$ for ${\bf q}$ along $\Gamma M$, lies between $\sim 0.1~\Gamma\Gamma=0.2~\Gamma M\approx 0.3\AA^{-1}$ and $\sim 0.2~\Gamma\Gamma=0.4~\Gamma M\approx 0.6\AA^{-1}$. At these wave vectors the energies of neutrons lie between $\sim 0.5 ~eV$ and $\sim 1~eV$ which are much easier for neutron scattering than $\sim 2~eV$ at the $M-$point. The typical shape of the peak intensity along $\Gamma K$ is similar to figure \ref{peak.fig}. Therefore according to this calculation, {\em the best chance of detecting spin-1 collective mode, is at the points between $\Gamma M$ and $\Gamma K$ which are closer to $\Gamma$, than $K$ or $M$.}  The BZ boundaries including $K$ and $M$ should be avoided.

Note that for these points, $I({\bq})$ becomes of the order of unity {\em and} the binding energy are typically $\hbar\omega_B\gtrsim 1~eV$. The larger $I({\bq})$ is, the easier will be exciting the triplet exciton. Equation (\ref{zandpsi.eqn}) of appendix A, explicitly shows the relationship between spatial extent of the excitonic wave-function and $Z({\bq})$. The small binding energies correspond to loosely bound particle-hole pairs and hence large wave-functions, which is according to (\ref{zandpsi.eqn}) synonymous to small peak intensities. Intuitively speaking, it becomes harder to excite larger objects by neutrons of wave-length $\lambda\sim\AA$.

\begin{figure}[t]
\begin{center}
\includegraphics[scale=0.8]{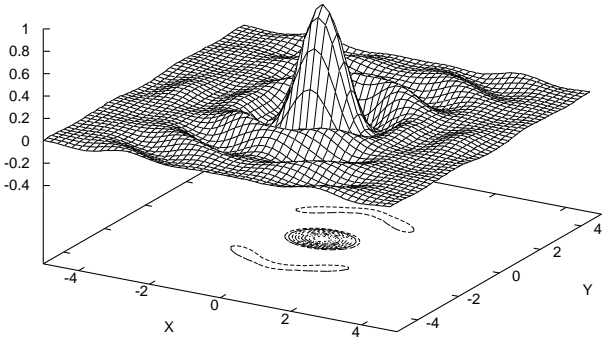}
\vspace{-0.5 cm}
\caption{The approximate bound state wave-function within the RPA approximation. The center of mass momentum ${\bq}$ is at the $M-$point. Unit of length is the lattice parameter $a\approx 2.46 \AA$. Contours corresponding to $|\psi_{\bq}({\br})|>0.1$ have been plotted in the base. Note the anisotropic pattern of contours. The normalization of the wave-function is such that $\psi_{\bq}(0)=1$. At this point the spatial extent of the bound state wave-function is not so large and it can be identified with triplet excitons. This state can not be excited so easily. Since it requires momentum transfers beyond the first BZ, at which atomic form factors wash out the neutron peak.}
\label{waveM.fig}
\end{center}
\end{figure}  

 Note that very close to $\Gamma-$point, where in figure \ref{peak.fig}, is not clear, according to (\ref{peak2.eqn}) $I({\bq})\approx Z({\bq})\sim q^3$  which corresponds to very large wave-functions. However as one expects form the qualitative behavior of binding energies (figure \ref{collective.fig}) and also the qualitative behavior of $I({\bq})$ (figure \ref{peak.fig}), as $q$ moves a bit away form $\Gamma-$point, the wave-function becomes of the order of a few lattice constants and hence visible by neutrons (figure \ref{waveGM0.2.fig}).

  In figures \ref{waveGM0.2.fig} and \ref{waveM.fig} the contours show the region in which $|\psi_{\bq}({\br})|>0.1$ which gives a good feeling about the spatial extent of the bound-state wave-function. They respectively correspond to $\bq=0.2\times {\vec {\Gamma M}}$ and $\bq= {\vec {\Gamma M}}$. The normalization of the wave-function is such that $\psi_{\bq}(\br=0)=1$. As can be seen form comparison of two figures, the spatial extent of the bound-state wave function for center of mass momentum corresponding to $M-$point, is of the order of two unit cells with highly anisotropic nature.

  Now let us discuss the effect of $\sigma$ bands. The $\sigma$ bands have their minima centered around the $\Gamma-$point. If we fit a quadratic dispersion to LDA data, we find that the inclusion of excitation form valence $\pi$ band to $\sigma$ band do not modify the collective mode qualitatively. The height of the window below the particle-hole continuum of figure \ref{ph_continuum.fig} along the $\Gamma K$ is not hight enough, so that the excitations to $\sigma$ do not shrink the window at all. However around $M-$point, at which the height of window is $2t$, the inclusion of excitations to $\sigma$ band, reduce the height by maximum amount of $\sim 0.5 t$ at the $M-$point. But as can be seen from figure \ref{collective.fig}, this does not open a decay channel for the normalized $U\sim 2t$.

\begin{figure}[t]
\vspace{1.0 cm}
\begin{center}
\includegraphics[scale=0.45]{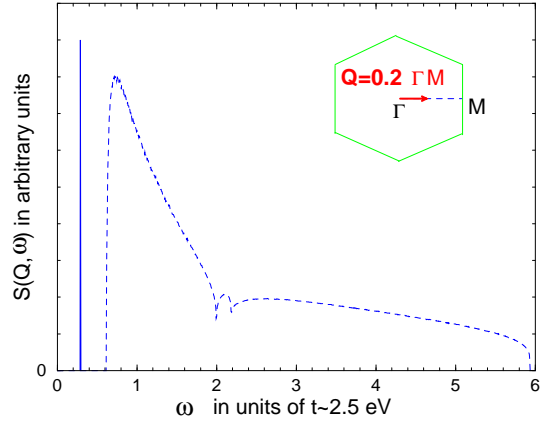}
\caption{Im$\chi^{RPA}(\omega)$ as a function of $\omega$ at ${\bf Q}=0.2\Gamma M\approx 0.3 \AA^{-1}$. Note that $\omega$ in this figure is in units of $t$. The particle-hole continuum is clearly seen in this figure (dashed line). Below the continuum at $\omega_s({\bf Q})\approx 0.30~t\approx 0.7~eV$ a very sharp peak in Im$\chi^{RPA}$ shows up (solid line).} 
\label{Sqw.fig}
\end{center}
\end{figure}  

\section{conclusion}
 We have evaluated the spin susceptibility of a graphene layer in RPA approximation for a short range interaction. This way we obtained a magnetic (spin-1) collective mode branch in non magnetic phase of graphite which exists in entire BZ with a very wide energy range form zero at $\Gamma$ and $K$ to $\sim 2~eV$ near the $M-$point. This branch is below the particle-hole continuum and is protected from Landau damping. Therefore it might provide a mechanism for transport of {\em spin-only currents}, over a wide energy range. We have shown that the long range tail of the Coulomb interaction of real graphite (3D stacks of graphene) does not destroy the conclusions drawn from the Hubbard interaction. We also presented the calculation of the weight of the neutron peaks. {\em The most appropriate region of the BZ to focus the neutron scattering experiments is region in between $\Gamma$ and $M$  and region mid-way between $\Gamma$ and $K-$points (closer to $\Gamma$ than $M$ or $K$). Hot or epithermal neutrons ($0.1-1.0~eV$) are more appropriate for this region of momentum transfers}. The dynamical form factor for a typical momentum transfer in the above region is schematically plotted in figure \ref{Sqw.fig}. At the above mentioned regions, in one hand one does not require very high energy neutrons. On the other hand,the binding is strong enough which leads to small enough wave-functions which can be excited with hot or epithermal neutrons.

\section{acknowledgment}
S. A. J. appreciates the Third World Academy of Sciences (TWAS) for 
financial support under TWAS fellowship for postdoctoral research and 
advanced training and the Institute of Mathematical 
Sciences (IMSc), chennai, India for hospitality. We thank Dr. Amir Murani
for encouraging discussions and correspondence on the possibility of study
of spin-1 collective modes in graphite by neutron scattering.

\appendix
\section{Appendix A: excitonic wave-function}
\label{wave-function.app}
  One can see more closely the relation between the $Z(\bq)$ and the binding energies and real space profile of the wave-function as follows. At the RPA approximation, the eigenvector of the exciton state with spin $S$ can be written as a linear combination of all the product eigenvectors of free charge carriers created in the crystal \cite{JSingh,Grosso}
\bearr
   \vert {\bq},S \rangle^{ex} = \sum_{\bk'} g_{\bq}({\bk'})~
   \vert {\bk'}+{\bq},-{\bk'};S \rangle
   \label{approx.exciton.eqn}
\eearr
where we have used the free electron-hole basis $\vert {\bk}+{\bq},-{\bk};S \rangle$ to expand the exciton wave-function. The triplet one which is relevant to our case is given by
\bearr
   \vert {\bk}+{\bq},-{\bk};1 \rangle &=&\frac{1}{\sqrt 3}
   \left( c^{c\dagger}_{{\bk}+{\bq}\up} d^{v\dagger}_{-{\bk}\up} 
   +c^{c\dagger}_{{\bk}+{\bq}\down} d^{v\dagger}_{-{\bk}\down}
   \right) | 0 \rangle \nn\\
   &+&\frac{1}{\sqrt 6} \left( 
   c^{c\dagger}_{{\bk}+{\bq}\up} d^{v\dagger}_{-{\bk}\down} +
   c^{c\dagger}_{{\bk}+{\bq}\down} d^{v\dagger}_{-{\bk}\up}
   \right) | 0 \rangle \nn
\eearr
where $d^{v\dagger}_{-{\bk}-\sigma}\equiv c^{v}_{{\bk}\sigma}$ creates a hole with spin projection $-\sigma$ and momentum $-{\bk}$. The exciton eigenvalue problem in free p-h basis becomes \cite{Grosso}
\bearr
  &&\left[ E^c({\bk}+{\bq})-E^v({\bk}) - W_{\bq}\right]g_{\bq}({\bk})\nn\\
  &+& \sum_{{\bk'}} U({\bk},{\bk'}) g_{\bq}({\bk'}) = 0
  \label{exciton.eqn}
\eearr
with kernel $U({\bk},{\bk'})$ given by
\bearr
   U({\bk},{\bk'}) &=& 
   -\langle c{\bk}+{\bq};v{\bk'}|U|c{\bk'}+{\bq};v{\bk} \rangle \nn\\
   &&+2\delta_{S,0}\langle c{\bk}+{\bq};v{\bk'}|U|v{\bk};c{\bk'}+{\bq}\rangle
   \label{kernel.eqn}
\eearr
Here $E^c({\bk}+{\bq})$ and $E^v({\bk})$ are {\em total} energies of single particle in conduction and valence band including their interaction energy with the other electrons of the band. The interaction between different pairs is included in in kernel $U({\bk},{\bk'})$ \cite{JSingh}. It can be seen from equation (\ref{kernel.eqn}) that in triplet channel kernel becomes a direct Coulomb term and moreover it is {\em attractive}. 

  The RPA approximation amounts to ignore the self-energy effects and write (i.e. solving a two-body problem)
\bearr
   E^c({\bk}+{\bq}) \to \varepsilon^c_{{\bk}+{\bq}/2} &=&
   \varepsilon_{{\bk}+{\bq}/2}\\
   E^v({\bk})       \to \varepsilon^v_{{\bk}-{\bq}/2} &=& 
   -\varepsilon_{{\bk}-{\bq}/2}
\eearr
and $   U({\bk},{\bk'})=-U$ for the Hubbard model. Note that wave-vectors are such that the wave-equation becomes manifestly time reversal invariant (${\bk} \to -{\bk}$), as can be seen from the wave-equation below
\bearr
   \left[\varepsilon_{{\bk}+{\bq}/2}+\varepsilon_{{\bk}-{\bq}/2}
   -\hbar\omega_s({\bq})\right]g_{\bq}({\bk})-U\sum_{{\bk'}}g_{\bq}({\bk'})=0
   \nn    \label{wave1.eqn}
\eearr
Let $\sum_{{\bk'}}g_{\bq}({\bk'})=C$, solve for $g_{\bq}({\bk})$ and sum over ${\bk}$ to obtain the self-consistency equation
\be
   C=\sum_{{\bk}}\frac{UC}{\varepsilon_{{\bk}+{\bq}/2}+
   \varepsilon_{{\bk}-{\bq}/2}-\hbar\omega_s({\bq})}\label{selfcons.eqn}
\ee
which is exactly The RPA approximation, equation (\ref{SZS.eqn}). The wave-function of the {\em single} particle-hole pair can be approximated by Fourier transform\footnote{Since the free particle-hole basis is composed of Bloch orbitals, therefore the Fourier transform is only an approximation} of $g_{\bq}({\bk})$. Thus we have
\bearr
  \psi_{\bq}(\br) &=&\frac{1}{N} \sum_{\bk}\frac{\exp(i{\bk}.{\br})}
  {\varepsilon_{{\bk}+{\bq}/2}+\varepsilon_{{\bk}-{\bq}/2}-
  \hbar\omega_s({\bq})} \label{psisum.eqn}\\
  &=&\frac{A}{4\pi^2}\int d^2{\bk}\frac{\exp(i{\bk}.{\br})}
  {\varepsilon_{{\bk}+{\bq}/2}+\varepsilon_{{\bk}-{\bq}/2}
  -\hbar\omega_s({\bq})} \label{psiint.eqn}
\eearr
with $A$, being the unit cell area, which shows that the wave function is purely real, as it should be. Comparing equations (\ref{zsum.eqn}) and (\ref{psisum.eqn}) shows that 
\be
  Z^{-1}({\bq})=\frac{U^2}{\pi} \sum_{\bk}\left|g_{\bq}({\bk}) \right|^2
  =\frac{U^2}{\pi}\int\frac{d^2{\br}}{A}\left|\psi_{\bq}({\br}) \right|^2
  \label{zandpsi.eqn}
\ee
Note that we leave $C$ in equation (\ref{selfcons.eqn}) is left undetermined. The normalization is such that $\psi_{\bq}({\br}=0)=1$. Equation (\ref{zandpsi.eqn}) shows that the intensity of collective mode peaks is a measure of inverse spatial extent of the particle-hole bound state. Smaller particle-hole bound states corresponds to sharper neutron scattering peaks.



\begin{thebibliography}{50}
\bibitem{kopelevich} Y. Kopelevich et al., cond-mat/0209442
\bibitem{Pauling} L. Pauling, {\em The Nature of The Chemical Bond}, 
  Cornell University Press (1960)
\bibitem{SZS} G. Baskaran and S. A. Jafari, Phys. Rev. Lett. {\bf 89}, 
   016402 (2002)
\bibitem{Jortner} J. Jortner, {\em et al.}, J. Chem. Phys. {\bf 42}, 309 (1965)
\bibitem{Goldoni} A. Goldoni, {\em et al.}, Synthetic Metalas, {\bf 77}, 
  189-94 (1996)
\bibitem{FLEX} N. E. Bickers and D. J. Scalapino, Ann. Phys. {\bf 193}, 
   206(1989)
\bibitem{Saitobook} R. Saito, G. Dresselhaus, and M. Dresselhaus, 
   {\it Physical Properties of Carbon Nanotubes}, Imperial College Press,
   London, 1998.
\bibitem{Wallace}
        P. R.~Wallace, Phys. Rev. {\bf 71}, 622 (1947); {\bf 72}, 258 (1947)
\bibitem{Diploma} S. A. Jafari, Diploma thesis, the Abdus Salam ICTP (2002)
\bibitem{paco} J. Gonz\'alez, F. Guinea, and M.A.H. Vozmediano, 
  Nucl.  Phys. B{\bf 424}, 595 (1994); Phys. Rev. Lett.,
  {\bf 77} 3859 (1996); cond-mat/0007337; D.V. Khveshchenko, cond-mat/0101306 
\bibitem{Murani} Amir Murani (private communication)
\bibitem{JSingh} J. Singh, {\em The dynamics of excitons}, in Solid State Physics, 
  H. Ehrenreich and D. Turnbull Eds, {\bf 38}, 295 (1984)
\bibitem{Grosso} G. Grosso and G. Pastori Parravichini, {\it Solid State Physics},
  Academic Press, 2000
\bibitem{Lovesey1} S. W. Lovesey, 
  {\it Condensed Matter Physics, Dynamic Correlations}, Benjamin/Cummings, 
  Reading, Massachusetts, 1980
\bibitem{Lovesey2} S. W. Lovesey, 
  {\it Theory of neutron scattering form condensed matter}, vol 1 and 2, 
  Clardon Press, Oxford, 1984
\bibitem{BJflex} G. Baskaran and S. A. Jafari (to be published)
\end{thebibliography}
\end{document}